\documentclass[aip,pop,reprint,showpacs,amsmath,amssymb]{revtex4-1}
\usepackage{graphicx}
\usepackage{bm}

\begin{document}


\title{Wake-induced bending of two-dimensional plasma crystals}
\author{T. B. R\"ocker}
\email{tbr@mpe.mpg.de}
\author{A. V. Ivlev}
\email{ivlev@mpe.mpg.de}
\author{S. K. Zhdanov}
\affiliation{Max Planck Institute for Extraterrestrial Physics, 85741 Garching, Germany}
\author{L. Cou\"edel}
\affiliation{CNRS, Aix-Marseille-Universit\'e, Laboratoire de Physique des Int\'eractions Ioniques et Mol\'eculaires, 13397
Marseille Cedex 20, France}
\author{G. E. Morfill}
\affiliation{Max Planck Institute for Extraterrestrial Physics, 85741 Garching, Germany}
\date{\today}

\begin{abstract}
It is shown that the wake-mediated interactions between microparticles in a two-dimensional plasma crystal affect the shape
of the monolayer, making it non-flat. The equilibrium shape is calculated for various distributions of the particle number
density in the monolayer. For typical experimental conditions, the levitation height of particles in the center of the
crystal can be noticeably smaller than at the periphery. It is suggested that the effect of wake-induced bending can be
utilized in experiments, to deduce important characteristics of the interparticle interaction.
\end{abstract}

\pacs{52.27.Lw, 52.27.Gr}

\maketitle

\section{Introduction}

Two-dimensional (2D) complex plasmas have been actively investigated in ground-based experiments since the two decades
\cite{Chu94,Thomas94,Hayashi94,Maddox94,Melzer94}. The focus has been made on experiments with crystalline and liquid
monolayers, representing an excellent natural model system for studies of generic phenomena occurring in classical 2D
liquids and solids \cite{Morfill2009,Ivlevbook}. Such monolayers are usually obtained in radio-frequency (rf) plasma
discharge chambers \cite{Thomas96,Konopka97,Samsonov99,Konopka00,KonopkaPhD,Nosenko06}, where the negatively charged
monodisperse microparticles levitate above the horizontal rf electrode due to the balance between the electrostatic force
(exerted in the electrode sheath via an inhomogeneous vertical electric field) and gravity.

The sheath field induces a strong vertical plasma flow toward the electrode, and each microparticle acts as the lens causing
the flowing ions to focus downstream from it. This results in the formation of plasma wakes ``attached'' to microparticles
\cite{Ishihara1997,Lampe2000,Melzer2000a,Hou2001,Vladimirov2003,Samarian2005,Miloch2010,Kompaneets2007,Ivlev05}. The wakes
exert attractive forces on the neighboring particles and make the pair interactions non-reciprocal. Nevertheless, the
collective behavior of such monolayers remains {\it exactly} conservative (i.e., apart from a frictional neutral-gas
damping, it can still be described by an effective Hamiltonian) until a certain threshold in the particle number density is
reached. This threshold depends on the strength of the vertical confinement and identifies the onset of the mode-coupling
instability (MCI) \cite{Ivlev03,Ivlev00_1,Zhdanov09,Couedel2010,Couedel11,Roecker14,Liu2010}. In the unstable regime the
particles acquire anomalous kinetic energy which is converted from the flowing plasma (due to resonance coupling between the
horizontal and vertical wave modes, mediated by non-reciprocal particle-wake interactions). If the MCI threshold is reached,
the instability can only be suppressed by frictional damping, e.g., by increasing the gas pressure.

\begin{figure}
\centering
\includegraphics[width=0.9\columnwidth,clip=]{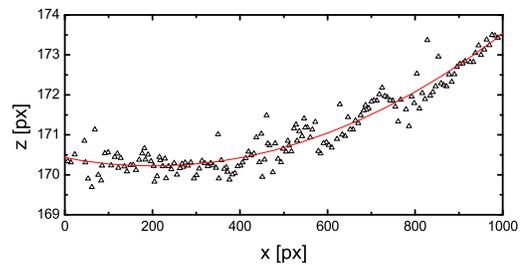}
\caption{\label{expobs} Illustration of bending of a 2D plasma crystal seen in experiments. A horizontal disk-shaped 2D
crystalline monolayer was formed by monodisperse melamine-formaldehyde particles of 9.19~$\mu$m diameter, levitated in an
argon discharge above a flat rf electrode of about $20$~cm diameter.
Particles were illuminated by a {\it vertical} thin ($\simeq2$ interparticle distances) laser sheet passing through
the center of the monolayer. Shown are the levitation heights of the traced individual particles in the illuminated region,
the solid line is a parabolic fit. The levitation height $z$ turns out to be dependent on the radial distance $x$ from the
center, where $z(x)$ attains a minimum (and where the areal density of particles is the highest). The pixel size is
$57~\mu$m, the vertical coordinate is measured from the electrode surface.}
\end{figure}

Theoretical studies of 2D complex plasmas usually assume that monolayers are flat, i.e., all particles levitate on the same
height, and that the number density is constant. In fact, the density normally has a maximum near the center (monolayers are
typically disk-shaped) and decreases towards the periphery. This inhomogeneity is caused by a weak (radially-dependent)
horizontal confinement, primarily due to edge effects in the discharge \cite{Quinn96,Zhdanov03,Sheridan09,Durniak10}. The
levitation height also varies across the monolayer -- it turns out to be the lowest at the center, so that the monolayer has
an observable curvature (``bending''), as illustrated in Fig. \ref{expobs}. To the best of our knowledge, the cause of the
bending has never been investigated, e.g., it may be associated with the edge effects too.

We would like to point out a mechanism of the bending which must always operate in experiments with 2D complex plasmas:
Along with the ``individual'' forces acting on each levitated particle (gravity, electrostatic, ion drag, thermophoretic,
etc.), there is also a ``collective'' contribution from the wakes of the neighboring particles. Unlike the direct
electrostatic interaction between charged particles, the wake-mediated interactions break the symmetry \cite{Steinberg01}.
The forces from the wakes of the neighboring particles are pointed downwards and, obviously, the resulting net force should
be dependent on the local particle density and configuration in the monolayer. Thus, the wake-mediated interactions should
always cause bending of the monolayer, and one could expect that the effect should be the strongest in the center, where the
particle density is the highest.

In this paper we analyze the effect of the wake-mediated interactions on bending of 2D plasma crystals. We show that, under
typical experimental conditions the magnitude of the effect can be quite significant: The difference between the levitation
height in the center and at the periphery can be as large as $\sim10\%$ of the mean interparticle distance. We suggest that
the bending phenomenon can be utilized in experiments, to deduce important characteristics of the interparticle interaction.

\section{Bending}
Let us consider a monolayer composed of $N$ identical charged particles whose equilibrium is determined by the balance
between the wake-mediated pair interactions and the forces of external confinement. The equilibrium configuration can be
described by the ``bending surface'' $z(x_i,y_i)$, where $\{x_i,y_i\}$ are the equilibrium horizontal coordinates of the
$i$th particle ($1\leq i\leq N$).

The horizontal confinement is usually poorly known, and therefore determining the horizontal particle distribution ``from
the first principles'' is hardly possible. On the other hand, the horizontal positions can be directly obtained from the
experimental top-view observations. Hence, one can use the measured coordinates $\{x_i,y_i\}$ as the input for calculating
the vertical displacement $z(x_i,y_i)$.

Parameters of the vertical confinement can be accurately measured in experiments with individual particles \cite{Ivlev00_2}.
Typically, the confinement is harmonic to a good accuracy. It is characterized by the vertical eigenfrequency $\Omega_{\rm
v}$; nonlinear terms can also be deduced and included in the analysis, for the sake of simplicity we neglect them here. For
the interparticle interactions we implement a simple Yukawa/point-wake model \cite{Couedel11,Rocker2012a}, which is
characterized by the negative particle charge $-Q$, effective screening length $\lambda$, positive wake charge $q$, and wake
length $\delta$. The results can be straightforwardly generalized for an arbitrary model of the wake potential.

\begin{figure}
\centering
\includegraphics[width=\columnwidth,clip=]{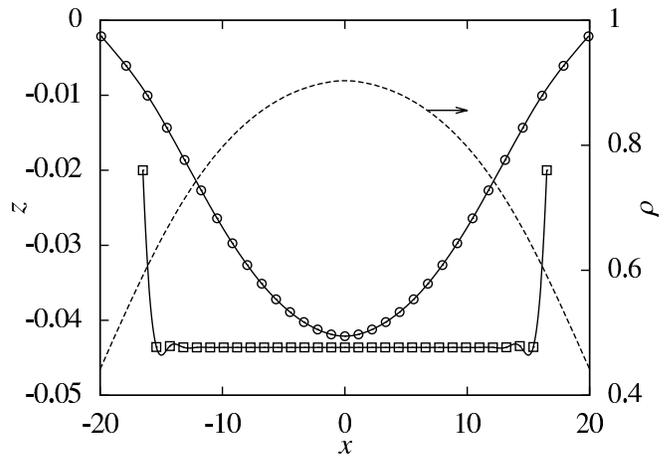}
\caption{\label{fig:1D} Bending of a 1D particle string (dimensionless units). The vertical displacement $z(x_i)$ of 31
particles is plotted (relative to the equilibrium height of a single particle). Two examples are shown, corresponding to a
constant (squares) and variable (circles) linear density $\rho$. The dashed line represents $\rho(x_i)$ for the latter case,
the constant-density value coincides with its maximum. The results are for the screening parameter $\kappa=1.1$
(at the maximum) and the wake dipole moment $\tilde{q} \tilde{\delta} = 0.24$, the eigenfrequency of vertical confinement
$\Omega_{\rm v} = 2.24$ corresponds to the MCI threshold (note that $z\propto\Omega_{\rm v}^{-2}$, see text for details).}
\end{figure}

In this section we normalize all lengths by $\lambda$, while the interaction potential and the vertical eigenfrequency are in units of the Debye energy scale
$Q^2/\lambda$ and the dust-lattice frequency $\sqrt{Q^2/M\lambda^3}$ , respectively
(where $M$ is the particle mass). The screening parameter $\kappa=\Delta/\lambda$ is defined for the mean interparticle
distance $\Delta$ at the {\it maximum} density. Furthermore, we introduce the normalized wake charge $\tilde{q} = q/Q$ and
length $\tilde{\delta} = \delta/\lambda$, so that the Yukawa/point-wake potential of a particle reads
\begin{equation}\label{ypw}
\phi({\bf r}) = \frac{e^{-r}}{r} -\tilde{q} \frac{e^{-r_{\delta}}}{r_{\delta}},
\end{equation}
where $r$ is the distance to the particle and $r_{\delta}=\sqrt{x^2+y^2+(z+\tilde{\delta})^2}$ is the distance to the wake.
The vertical displacement $z_i\equiv z(x_i,y_i)$ of the $i$th particle is governed by the following equation readily derived
from the vertical force balance:
\begin{equation}\label{eq:1D}
	\sum_{j \neq i} \frac{\partial \phi}{\partial z} \Big|_{{\bf r}_j-{\bf r}_i} + \Omega^2_{\rm v} z_i = 0,
\end{equation}
where the summation is over the neighbors. Thus, the bending surface is generally determined by the solution of $N$ coupled
equations (\ref{eq:1D}).

In order to demonstrate the generic properties of bending and, simultaneously, to highlight the important effect of
neighbors, we shall start the consideration with a 1D particle string and then discuss 2D monolayers.

Figure~\ref{fig:1D} shows the ``bending lines'' $z(x_i)$ calculated from Eq.~(\ref{eq:1D}) for a 1D string comprised of
$N=31$ particles. We present results for a constant density (squares) and variable density (circles). In the latter case the
horizontal particle positions were deduced from molecular dynamics simulations (described in Refs. \cite{Roeckerthesis,Zhdanov03}). The resulting density distribution (dashed line) has a maximum at the center, where it is equal to the
value used for the constant-density case.

Remarkably, for a constant-density case the string remains flat, i.e., practically all particles except those at the ends
are shifted downwards as the whole. This indicates that the magnitude of the shift at a given particle density is mostly
determined by the nearest neighbors, i.e., even for relatively small ($N\geq5-6$) clusters with equal spacing the vertical
displacement is practically independent of $N$. In the case of variable density, the net force from the mutual interactions
rapidly decreases toward the ends and, therefore, $|z_i|$ decreases as well, but the displacement at the center is
practically the same as for the constant-density case.

\begin{figure}
\centering
\includegraphics[width=\columnwidth]{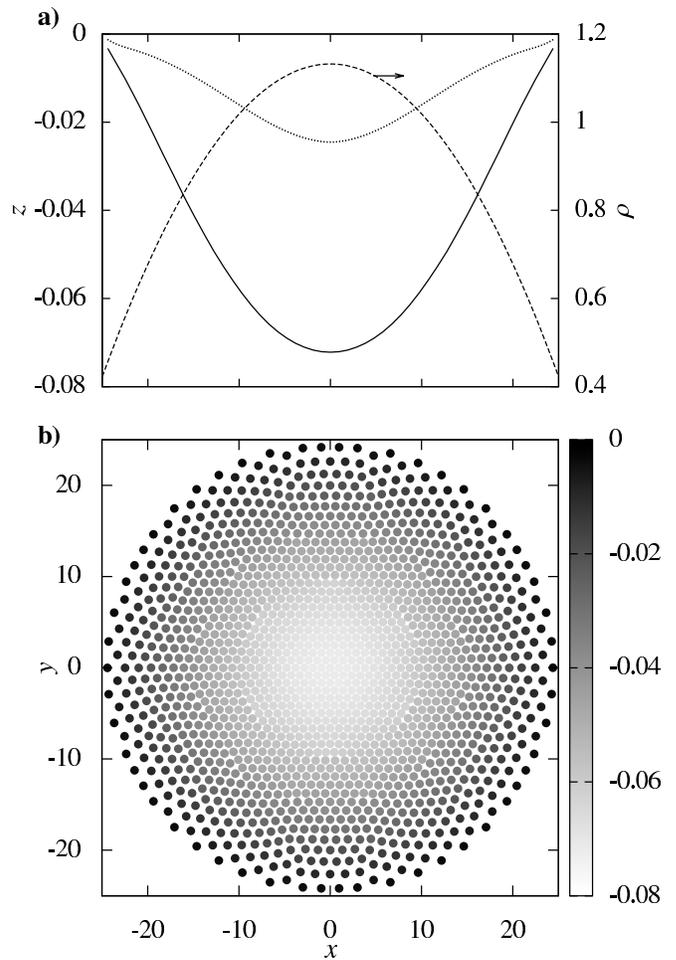}
\caption{\label{fig:2D} Bending of a 2D crystalline monolayer (dimensionless units). The crystal has a form of an
axially-symmetric disc composed of 1519 particles, the radial dependence of the (average) areal density $\rho$ is shown in
(a) by the dashed line. The corresponding radial dependence of the vertical displacement $z$ is depicted by the solid line,
and the bending line for a 1D particle string with the same linear density distribution is also shown for comparison (dotted
line). The top view of the bent monolayer is presented in (b), where the magnitude of the local vertical displacement is
grayscale-coded. The results are for the screening parameter $\kappa = 0.91$ (at maximum density) and the wake dipole moment
$\tilde{q} \tilde{\delta}=0.24$, the eigenfrequency of vertical confinement $\Omega_{\rm v}=4.22$ corresponds to the MCI
threshold.}
\end{figure}

Figure~\ref{fig:2D} illustrates bending of a 2D monolayer. The radial dependence of the vertical displacement is presented
in Fig.~\ref{fig:2D}a (solid line) for a given areal density distribution (dashed line), and the complementary results for a
1D string (dotted line, with the same linear density distribution) are also shown. Obviously, bending of monolayers and
strings is qualitatively the same, but its magnitude $H$ (maximum vertical displacement) is about 3 times larger in the 2D
case ($H \simeq 0.08$). So, the magnitude of $H$ for a given density appears to be primarily determined by the lattice
coordination number (whose ratio for 2D and 1D lattices is 3). Furthermore, Fig.~\ref{fig:2D}b demonstrates that near the
monolayers center, where the density is practically constant, the particles levitate at almost the same height.

\begin{figure}[t]
\centering
\includegraphics[width=0.95\columnwidth]{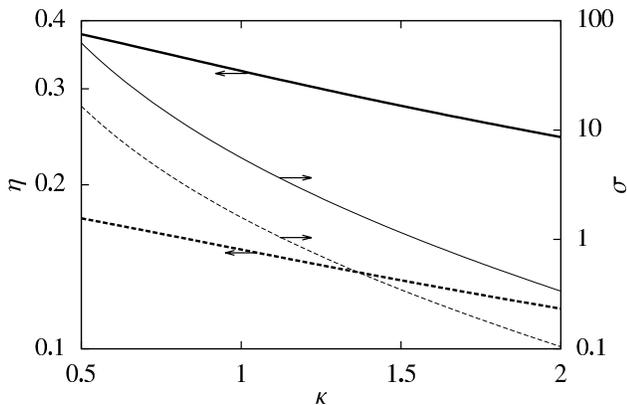}
\caption{\label{kapcurve} Functions $\sigma(\kappa)$ (thin lines) and $\eta(\kappa)$ (thick lines) determining the universal
dependencies for the bending magnitudes in Eqs.~(\ref{Hform}) and (\ref{Hcrit}), respectively. The shown curves are for 2D
monolayers (solid lines) and 1D strings (dashed lines), the results are obtained in the limit $\delta^2 \ll 1$.}
\end{figure}

A general dependence of the bending magnitude $H$ on the parameters of pair interaction and confinement can be derived from
the following simple consideration: The bending is the result of interaction asymmetry due to the wake term in
Eq.~(\ref{ypw}). For the latter, the dependence on $z$ is negligible when $\tilde\delta|dz/dr_{\rm h}| \ll\kappa$ (where
$r_{\rm h}$ is the horizontal coordinate). As one can see from Fig.~\ref{fig:2D}, $|dz/dr_{\rm h}|$ is typically very small
($\sim10^{-2}$), so this condition is always satisfied. Then the first term in Eq.~(\ref{eq:1D}) does not depend on $z_i$
and we get
\begin{equation}\label{Hform}
 H = \frac{\tilde{q}\tilde{\delta}}{\Omega^2_{\rm v}} \sum_{i \neq 0} \frac{1+\xi_i}{\xi^3_i}e^{-\xi_i}
 \equiv \frac{\tilde{q}\tilde{\delta}}{\Omega^2_{\rm v}} \sigma(\kappa,\tilde\delta)
\end{equation}
where $\xi_i = \sqrt{r_{{\rm h}i}^2+\tilde{\delta}^2}$ is the distance to the wake of the $i$th neighbor. We immediately
conclude that the bending magnitude is directly proportional to the effective wake charge and inversely proportional to the
squared confinement frequency. One can further simplify the obtained expression by noting that the dependence of $\sigma$ on
$\tilde\delta$ can be neglected when $\tilde{\delta}^2 \ll\kappa^2$. Typically, the screening parameter $\kappa$ is about
unity, so that for $\tilde{\delta} \lesssim 0.3$ the distance is reduced to $\xi_i \simeq r_{{\rm h}i}$. Hence, in practice
$\sigma$ is a (strongly decreasing) function of $\kappa$ only, as shown in Fig. \ref{kapcurve} (thin lines), and $H$ is
proportional to the wake dipole moment $\tilde{q} \tilde{\delta}$. We note that the functions representing 2D and 1D cases
in Fig.~\ref{kapcurve} (solid and dashed lines, respectively) are similar and their ratio also approaches 3 for
$\kappa\gtrsim1$ (cf. bending profiles in Fig.~\ref{fig:2D}a).

The strongest {\it equilibrium} bending is observed when $\Omega_{\rm v}$ is reduced down to the critical (threshold)
confinement frequency $\Omega_{\rm cr}$ corresponding to the onset of MCI. For $\tilde\delta^2\ll1$, the combination
$(1-\tilde{q})^{-1}\Omega^{2}_{\rm cr}$ is a function of $\kappa$ only \cite{Couedel11}, and this rather strong dependence
turns out to be very similar to $\sigma(\kappa)$. Therefore, the maximum equilibrium magnitude of the bending is given by
the following simple dependence:
\begin{equation}\label{Hcrit}
 H_{\rm cr} = \tilde{q}\tilde{\delta} \frac{\sigma(\kappa)}{\Omega^2_{\rm cr}(\kappa)} \equiv \frac{\tilde{q}
 \tilde{\delta}}{1-\tilde{q}} \eta(\kappa)
\end{equation}
where $\eta(\kappa)$ is a relatively weak function shown in Fig.~\ref{kapcurve} (thick lines).

\section{Conclusion}

Bending of 2D plasma crystals is not only an interesting effect caused by the interaction non-reciprocity. We showed that
its magnitude can be quite large, about one tenth of the effective screening length, which gives about $40-80~\mu$m for
typical experiments with 2D complex plasmas \cite{Couedel11}. This value is comparable with the thickness of the horizontal
laser sheets used in experiments to illuminate levitated particles. Therefore, we suggest that the bending phenomenon can be
utilized to deduce important characteristics of the interparticle interaction.

The only well-established experimental method to determine the screening length and the {\it effective} particle charge,
$\sqrt{1-\tilde{q}}\,Q$ \cite{Couedel11}, is based on fitting theoretical dispersion relations to experimental fluctuation
spectra \cite{Couedel2010,Nunomura02}. Another recently proposed \cite{Rocker2012a} method of mapping self-consistent wake models to the
effective dipole moment $\tilde q\tilde\delta$ allows us to relate the wake parameters to the screening parameter $\kappa$.
However, the accuracy of this method has not been tested so far.

The results presented in this paper allow us to identify at least two simple algorithms to obtain certain combinations of
the interaction parameters. For this, let us rewrite Eqs.~(\ref{Hform}) and (\ref{Hcrit}) in the {\it dimensional form}:

$\bullet$ From Eq.~(\ref{Hcrit}) we derive
\begin{equation}\label{Qeff}
	\frac{q\delta}{Q-q}=H_{\rm cr}\eta^{-1}(\kappa). \nonumber
\end{equation}
By using the value of $\kappa$ in the center of the monolayer (e.g., deduced from fitting of fluctuation spectra) as the
input parameter and measuring the bending magnitude $H_{\rm cr}$ at the onset of MCI, we obtain a combination of $Q$, $q$,
and $\delta$. Since $\eta$ only weakly depends on $\kappa$, this algorithm provides fairly accurate results even if the
accuracy of measuring $\kappa$ is poor.

$\bullet$ From Eq.~(\ref{Hform}) we get
\begin{equation}
	Q q \delta=(2\pi f_{\rm v})^2MH\lambda^3\sigma^{-1}(\kappa). \nonumber
\end{equation}
In this case, it is sufficient to measure $H$ at a given resonance frequency of a single particle, $f_{\rm v}$ (in units of
Hz), and also use $\lambda$ as the input parameter, which yields another combination of $Q$, $q$, and $\delta$. Since
$\sigma(\kappa)$ is a strong function, this algorithm requires very accurate determination of $\lambda$. Note that one can
divide the two obtained combinations to directly deduce the squared effective charge $(Q-q)Q$.

We conclude that the bending phenomenon commonly observed in experiments with 2D complex plasmas can provide us with a
simple powerful method to deduce important characteristics of the interparticle interaction. On the other hand, of course,
one should keep in mind that the bending could also be associated with other phenomena, e.g., with a possible inhomogeneity
of the sheath electric field and/or particle charge in the horizontal direction. The magnitude (and even the sign) of such
effects are unknown and can only be measured in dedicated experimental tests.

\begin{acknowledgments}
We appreciate funding from the European Research Council under the European Unions Seventh Framework Programme/ERC Grant
Agreement 267499, and from the French-German PHC PROCOPE Program/Project 28444XH/55926142. T.B.R. acknowledges Alexandra
Heimisch for the helpful support.
\end{acknowledgments}


\begin{thebibliography}{99}

\bibitem{Chu94} J.H. Chu and Lin I, Phys. Rev. Lett. {\bf 72}, 4009 (1994).

\bibitem{Thomas94} H. Thomas, G.E. Morfill, V. Demmel, J. Gorree, B. Feuerbacher, and D. M\"ohlmann, Phys. Rev. Lett. {\bf 73}, 652 (1994).

\bibitem{Maddox94} J. Maddox, Nature {\bf 370}, 411 (1994).

\bibitem{Hayashi94} Y. Hayashi and K. Tachibana, Jpn. J. Appl. Phys. {\bf 33}, L804 (1994).

\bibitem{Melzer94} A. Melzer, T. Trottenberg, and A. Piel, Phys. Lett. A {\bf 191}, 301 (1994).

\bibitem{Morfill2009} G.E. Morfill and A.V. Ivlev, Rev. Mod. Phys. {\bf 81}, 1353 (2009).

\bibitem{Ivlevbook} A. Ivlev, H. L\"owen, G. Morfill, and C.P. Royall, \textit{Complex Plasmas and Colloidal Dispersions:
    Particle-Resolved Studies of Classical Liquids and Solids}, World Scientific, Singapore (2012).

\bibitem{KonopkaPhD} U. Konopka, \emph{Wechselwirkungen geladener Staubteilchen in Hochfrequenzplasmen}, Phd thesis (2000).

\bibitem{Thomas96} H. Thomas and G. Morfill, Nature {\bf 379}, 806 (1996).

\bibitem{Konopka97} U. Konopka, L. Ratke, and H. M. Thomas, Phys. Rev. Lett. {\bf 79}, 1269 (1997).

\bibitem{Samsonov99} D. Samsonov, J. Goree, Z.W. Ma, A. Bhattacharjee, H.M. Thomas, and G.E. Morfill, Phys. Rev. Lett. {\bf 83}, 3649 (1999).

\bibitem{Konopka00} U. Konopka, G.E. Morfill, and L. Ratke, Phys. Rev. Lett. {\bf 84}, 891 (2000).

\bibitem{Nosenko06} V. Nosenko, J. Goree, and A. Piel, Phys. Plasmas {\bf 13}, 032106 (2006).

\bibitem{Ishihara1997} O. Ishihara and S.V. Vladimirov, Phys. Plasmas \textbf{4}, 69 (1997).

\bibitem{Lampe2000} M. Lampe, G. Joyce, G. Ganguli, and V. Gavrishchaka, Phys. Plasmas \textbf{7}, 3851 (2000).

\bibitem{Melzer2000a} A. Melzer, V.A. Schweigert, and A. Piel, Physica Scripta \textbf{61}, 494 (2000).

\bibitem{Hou2001} L.-J. Hou, Y.-N. Wang, and Z.L. Miskovic, Phys. Rev. E {\bf 64}, 46406 (2001).

\bibitem{Vladimirov2003} S.V. Vladimirov, S.A. Maiorov, and O. Ishihara, Phys. Plasmas \textbf{10}, 3867 (2003).

\bibitem{Samarian2005} A. Samarian, S. Vladimirov, and B. James, JETP Letters \textbf{82}, 758 (2005).

\bibitem{Miloch2010} W.J. Miloch, Plasma Physics and Controlled Fusion \textbf{52}, 124004 (2010).

\bibitem{Kompaneets2007} R. Kompaneets, U. Konopka, A.V. Ivlev, V. Tsytovich, and G. Morfill, Phys. Plasmas \textbf{14}, 052108 (2007).

\bibitem{Ivlev05} A.V. Ivlev, S.K. Zhdanov and G.E. Morfill, Phys. Rev. E {\bf 71}, 016405 (2005).

\bibitem{Ivlev00_1} A.V. Ivlev and G.E. Morfill, Phys. Rev. E {\bf 63}, 016409 (2000).

\bibitem{Liu2010} B. Liu, J. Goree, and Y. Feng, Phys. Rev. Lett. \textbf{105}, 085004 (2010); B. Liu, J. Goree, and Y.
    Feng, Phys. Rev. Lett. \textbf{105}, 269901 (2010).

\bibitem{Ivlev03} A.V. Ivlev, U. Konopka, and G.E. Morfill, Phys. Rev. E {\bf 68}, 026405 (2003).

\bibitem{Zhdanov09} S.K. Zhdanov, A.V. Ivlev, and G.E. Morfill, Phys. Plasmas {\bf 16}, 083706 (2009).

\bibitem{Couedel2010} L. Cou\"edel, V. Nosenko, A.V. Ivlev, S.K. Zhdanov, H.M. Thomas, and G.E. Morfill, Phys. Rev. Lett. \textbf{104}, 195001 (2010).

\bibitem{Couedel11} L. Cou\"edel, S.K. Zhdanov, A.V. Ivlev, V. Nosenko, H.M. Thomas, and G.E. Morfill, Phys. Plasmas {\bf 18}, 083707 (2011).

\bibitem{Roecker14} T.B. R\"{o}cker, A.V. Ivlev, S.K. Zhdanov, and G.E. Morfill, Phys. Rev. E \textbf{89}, 013104 (2014).

\bibitem{Zhdanov03} S. Zhdanov, R.A. Quinn, D. Samsonov and G.E. Morfill, N. Journal Phys. {\bf 5}, 74 (2003).

\bibitem{Quinn96} R.A. Quinn, C. Cui, J. Goree and J.B. Pieper, Phys. Rev. E {\bf 53}, R2049 (1996).

\bibitem{Sheridan09} T.E. Sheridan, Phys. Plasmas {\bf 16}, 083705 (2009).

\bibitem{Durniak10} C. Durniak, D. Samsonov, N.P. Oxtoby, J.P. Ralph and S. Zhdanov, IEEE Trans. Plasma Sci. {\bf 38}, 2412 (2010).

\bibitem{Steinberg01} V. Steinberg, R. S\"utterlin, A. V. Ivlev, and G. Morfill, Phys. Rev. Lett. {\bf 86}, 4540 (2001).

\bibitem{Ivlev00_2} A. V. Ivlev, R. S\"utterlin, V. Steinberg, M. Zuzic, and G. Morfill, Phys. Rev. Lett. {\bf 85}, 4060 (2000).

\bibitem{Rocker2012a} T.B. R\"{o}cker, S.K. Zhdanov, A.V. Ivlev, M. Lampe, G. Joyce, and G.E. Morfill, Phys. Plasmas \textbf{19}, 073708 (2012).

\bibitem{Roeckerthesis} T.B. R\"ocker, \emph{Mode-coupling regimes in 2D plasma crystals}, Phd thesis (to be published).

\bibitem{Nunomura02} S. Nunomura, J. Goree, S. Hu, X. Wang, A. Bhattacharjee, and K. Avinash, Phys. Rev. Lett. {\bf 89}, 035001 (2002).

\end{thebibliography}
\end{document}